\documentclass[a4paper]{article}

\usepackage{amsmath,amsfonts,amssymb}
\usepackage{amsthm}
\usepackage{enumerate}
\usepackage{bbm}
\usepackage{graphicx}
\usepackage{lscape,epsf}
\usepackage[vcentermath]{youngtab}
\usepackage{rotating}
\usepackage{hyperref}

\def\ds{\displaystyle}
\def\bea{\begin{array}{c}}
\def\ea{\end{array}}
\def\be{\begin{equation}\bea\ds}
\def\ee{\ea\end{equation}}
\def\bee{\begin{equation}\begin{array}{rcl}\ds}
\def\eee{\end{array}\end{equation}}

\def\tr{{\rm Tr}\,}

\def\Hc{{\mathcal{H}}}

\def\tr{{\rm tr}\,}
\def\Tr{{\rm Tr}\,}

\author{D.~Melnikov$^{a,b}$, A.~Mironov$^{b,c,d}$, S.~Mironov$^{b,e}$, A.~Morozov$^{b,d}$ and An.~Morozov$^{b,d,f}$}

\title{\bf From Topological to Quantum Entanglement}

\begin{document}

\maketitle

\vspace{-6cm}
\begin{flushright}
ITEP/TH-25/18 \\
FIAN/TD-17/18\\
IITP/TH-15/18
\end{flushright}
\vspace{3cm}

\vspace{6pt}
\begin{center}
\textit{\small $^a$ International Institute of Physics, Federal University of Rio Grande do Norte, \\ Campus Universit\'ario, Lagoa Nova, Natal-RN  59078-970, Brazil}
\\ \vspace{6pt}
\textit{\small $^b$  Institute for Theoretical and Experimental Physics, B.~Cheremushkinskaya 25, Moscow 117259, Russia}\\ \vspace{6pt}
\textit{\small $^c$  I.~E.~Tamm Theory Department, Lebedev Physics Institute, Leninsky pr.~53, Moscow 119991, Russia}\\ \vspace{6pt}
\textit{\small $^d$  Institute for Information Transmission Problems,
Bolshoy Karetny per.~19 build.~1, Moscow 127051, Russia}
\\ \vspace{6pt}
\textit{\small $^e$  Institute for Nuclear Research of the Russian Academy of Sciences, \\ 60th October Anniversary prosp~7a, 117312 Moscow, Russia}\\ \vspace{6pt}
\textit{\small $^f$  Moscow Institute of Physics and Technology, Dolgoprudny 141701, Russia}
\\ \vspace{1cm}
\end{center}

\begin{abstract}
Entanglement is a special feature of the quantum world that reflects the existence of subtle, often non-local, correlations between local degrees of freedom. In topological theories such non-local correlations can be given a very intuitive interpretation: quantum entanglement of subsystems means that there are ``strings'' connecting them. More generally, an entangled state, or similarly, the density matrix of a mixed state, can be represented by cobordisms of topological spaces. Using a formal mathematical definition of TQFT we construct basic examples of entangled states and compute their von Neumann entropy.
\end{abstract}

\section{Introduction}

Quantum entanglement is a signature characteristics of the quantum world. As such it has been at the core of the discussion about the fundamentals of quantum physics for nearly a century~\cite{Review}. One particularly interesting feature of quantum entanglement is its non-locality, the existence of quantum correlations beyond the limits prescribed, for example, by naive causality~\cite{EPR}. The non-local correlations turn out to be a source of many tantalizing prospects for practical applications, including quantum metrology, fault-tolerant quantum computing, quantum cryptography and data protection. They reveal the intrinsic non-locality of quantum theory and emphasize, on the very basic level, the necessity to consider non-local models of quantum field theory and its uplifting to a full-fledged string theory~\cite{Morozov:1992ea}.

At the same time, there are systems in physics which do not have local degrees of freedom at all. Such systems are commonly called topological if their non-local degrees of freedom are only in topological, geometry independent, relation with each other. A typical example is the quantum Hall system -- a gapped state, whose non-local excitations (quasiparticles) only possess statistical interactions. Due to their non-locality, topological systems should  provide special instances of quantum entanglement, illuminating its non-local properties -- one can expect them to be best suited for the study of entanglement and its applications. One can even wonder if the entire subject of quantum entanglement is reduced to the topological entanglement,
and if the latter one can be treated in fully classical terms.

One typical example of the role of non-local correlations in quantum entanglement was considered in Refs.~\cite{Kitaev:2005dm,Levin:2006zz}. It was shown that the scale independent piece of the von Neumann entropy in $2+1$D theories is topological, in the sense that it does not depend on the geometry of the partition. It was argued that this piece matches the full entropy in some topological theories. The calculation of the entropy in such a setup was later elaborated in Ref.~\cite{Dong:08}, where it was noticed that in topological quantum field theories (TQFTs) von Neumann and R\'enyi entropies can be computed using the replica trick. These entropies are expressed in terms of topological invariants of links in different three-dimensional manifolds. Connections with links was recently revisited by the authors of Refs.~\cite{Balasubramanian:2016sro,Balasubramanian:2018por}, who associated von Neumann entropies to the links themselves. In particular, the Hopf link was compared with a maximally entangled EPR pair, while the generic multicomponent torus links were shown to be counterparts of the so-called GHZ states.

In this paper we further elaborate on the ideas discussed in~\cite{Balasubramanian:2016sro} and propose a different, more general, perspective on the connection between entanglement and topological quantum field theories. The latter possess a very elegant and intuitive mathematical description in terms of (co)bordisms, or ``time evolutions'' of closed manifolds. Cobordisms suggest a natural definition of quantum entanglement in TQFTs: the two subsystems are entangled if they are (co)bordant. Similarly, a cobordism of two equivalent manifolds is a time evolution, or the density matrix of a generically mixed state. In particular, partial tracing of the density matrix of a pure entangled state produces a mixed density matrix in a very illustrative way.

We demonstrate these properties on a set of examples complementary to those considered in Refs.~\cite{Dong:08,Balasubramanian:2016sro,Balasubramanian:2018por,Dwivedi:2017rnj}. We show that in some cases, to be entangling, the cobordisms need to be endowed with an additional structure supported by Wilson lines. Altogether this description gives a graphical idea of quantum entanglement between two subsystems: they need to be tied by strings of Wilson line operators. Somewhat counterintuitively, additional tangling of the strings may actually decrease the amount of entanglement. This was noticed also in Ref.~\cite{Balasubramanian:2016sro}, where the maximal two-partite entanglement was demonstrated by the simplest link -- the Hopf link.

The present authors' motivation for this work came from the study of the topological quantum computer~\cite{TQC}. In Ref.~\cite{topquco} we presented a reflection on the current status of knot theory from the point of view of possible quantum computing applications. One of interesting problems, which we believe has not so far been well elucidated in the literature is the interpretation of knot and link polynomials as quantum algorithms. We believe that the due analysis of the quantum code spaces and entanglement is a good starting point in this direction. See the review of L.~Kauffman~\cite{Kauffman:2013} for a nice summary of the necessary material and a more recent discussion on similar ideas, and appropriate references, in Ref.~\cite{Kauffman:2016}.

This work is organized as follows. In section~\ref{sec:TQFT} we remind a useful mathematical definition of TQFTs due to E.~Witten and M.~Atiyah. We make the analogy with ordinary quantum mechanics. We introduce the notion of quantum entanglement in section~\ref{sec:EE} and compute von Neumann entropy for a generic bipartite system in TQFT terms. In section~\ref{sec:examples} we consider some examples and  refine the topological condition for entanglement. We give brief conclusions and perspectives in section~\ref{sec:conclusions}.

\section{Topological quantum field theories}
\label{sec:TQFT}

Topological quantum field theories provide an interesting class of field theories with no local propagating degrees of freedom. In contrast with normal QFTs, the Hilbert space of these theories is often finite dimensional, which actually makes them closer to quantum mechanics, rather than field theory. Being halfway between quantum mechanics and quantum field theory TQFTs can help understanding, how quantum mechanical concepts generalize to QFTs.

An elegant mathematical definition of TQFT was given by E.~Witten~\cite{Witten:1988hf} and M.~Atiyah~\cite{TQFT-Atiyah} in terms of a functor (map) $Z$ between a category of topological spaces and a category of vector spaces. Sparing all the formal details of the definition we simply mention the main points and consequences:

\begin{itemize}
\item For any $D$-dimensional oriented space $\Sigma$, we associate a vector space $V=Z(\Sigma)$. In the explicit examples we will assume $D=2$, and examples of $\Sigma$ could be looked among Riemann surfaces, with some points possibly removed. In our further diagrammatic presentation we depict $\Sigma$ by a circle
$$Z\left(
\begin{array}{c}
\includegraphics[width=0.05\linewidth]{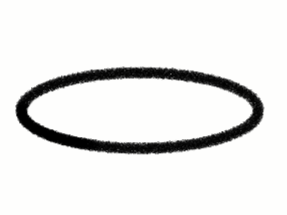}
\end{array}
\right) \ = \ \Hc\,.$$
Here the image of $\Sigma$ is a linear space $\Hc$, which we understand as a (local) Hilbert space. Consequently, we may consider a disjoint set of circles, which should be mapped to a direct product of Hilbert spaces
$$  Z(\Sigma_1\cup \Sigma_2\cup \cdots \cup\Sigma_n) \ = \ \Hc_1\otimes \Hc_2\otimes \cdots \otimes \Hc_n\,.$$

\item For a $D$-dimensional $\Sigma$ and $D+1$-dimensional manifold $M$, such that $\partial M = \Sigma$, we associate a vector $v=Z(M)$ in $V=Z(\Sigma)$. In our diagrammatic notations we say
$$Z\left(
\begin{array}{c}
\includegraphics[width=0.05\linewidth]{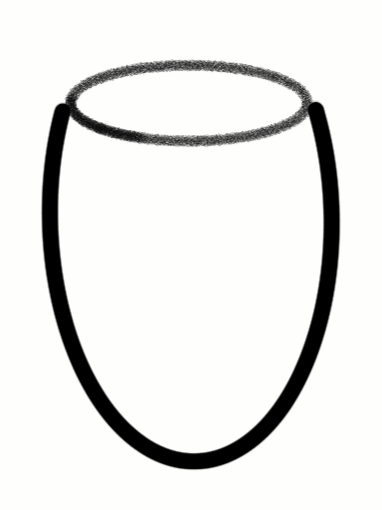}
\end{array}
\right) \ = \ |\psi\rangle \ \in \ \Hc\,.$$
That is a disk, whose boundary is a circle represents a state in the Hilbert space. However, since here we mainly think of $\Sigma$ being two-dimensional, $M$ will mostly be a three-dimensional space.

\item If $M$ is a closed manifold, then it can be understood as a manifold with an empty boundary. In such a case we associate the image of $M$ with a $C$-number: $Z(M)\in C$. Moreover, since manifolds $\Sigma$ are considered to be orientable, we associate with $\bar{\Sigma}$, which differs from $\Sigma$ by the choice of orientation, a dual vector space $Z(\bar\Sigma)=V^\ast$. This allows us to define the scalar product in terms of topological spaces. In particular, we identify
$$Z\left(
\begin{array}{c}
\includegraphics[width=0.1\linewidth]{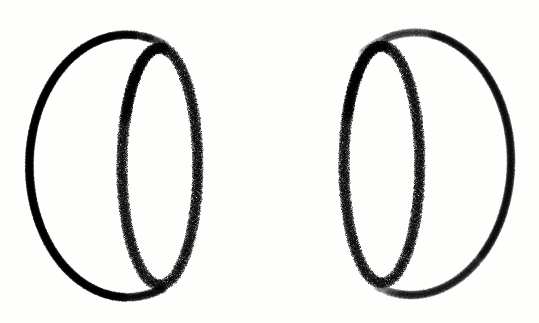}
\end{array}
\right)
\ = \
Z\left(
\begin{array}{c}
\includegraphics[width=0.06\linewidth]{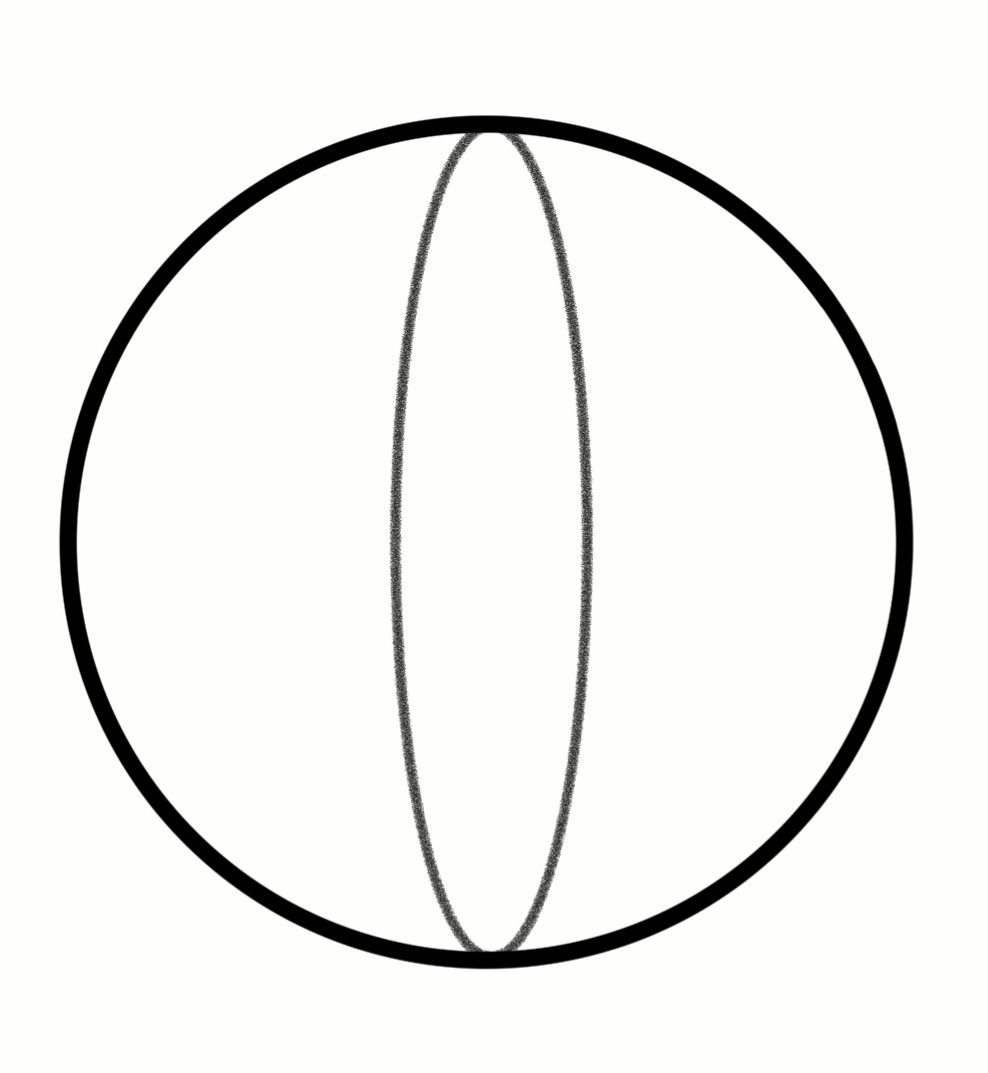}
\end{array}
\right)
 \ = \ \langle \psi | \chi \rangle \ \in C\,.$$
In other words, one can glue the topological spaces along boundaries, which differ by orientation, to form new topological spaces.

\item Cobordisms, or $D+1$-dimensional manifolds $M$ with two boundaries $\Sigma_1$ and $\Sigma_2$, with the appropriate choice of orientation, are mapped to linear maps: $Z(M): Z(\Sigma_1)\to Z(\Sigma_2)$. Using the gluing procedure discussed above, one can construct compositions of linear maps. States in the above example can be understood as cobordisms of a given $\Sigma$ to an empty manifold.
\end{itemize}

One can view cobordisms as evolutions of manifolds $\Sigma$. We note that $\Sigma$ can have a number of points removed. This is equivalent to introducing boundaries to $\Sigma$ themselves. Consequently, one should track the evolution of those boundaries, or points. We will come back to the discussion of such marked points when we consider some explicit examples.

Now we have defined some quantum mechanics (Hilbert space, states, scalar product and operators) in terms of topological spaces and cobordisms.\footnote{See Refs.~\cite{Dedushenko:2018aox,Dedushenko:2018tgx} for recent applications of TQFT quantum mechanics in supersymmetric theories.} For an explicit construction of such a map (functor) self-consistency relations need to be satisfied. The first example was constructed by E.~Witten in~\cite{Witten:1988hf}, where $Z$ was the functional integral of a non-Abelian Chern-Simons theory in three dimensions.

\section{Entanglement in TQFT}
\label{sec:EE}

TQFTs suggest quite an intuitive notion of the entanglement of two subsystems. Let us partition the system into a disjoint union $\Sigma_1\cup \Sigma_2$. Based on two circles, one can construct two basic cobordisms representing possible quantum states:
\be
\begin{array}{c}
\includegraphics[width=0.1\linewidth]{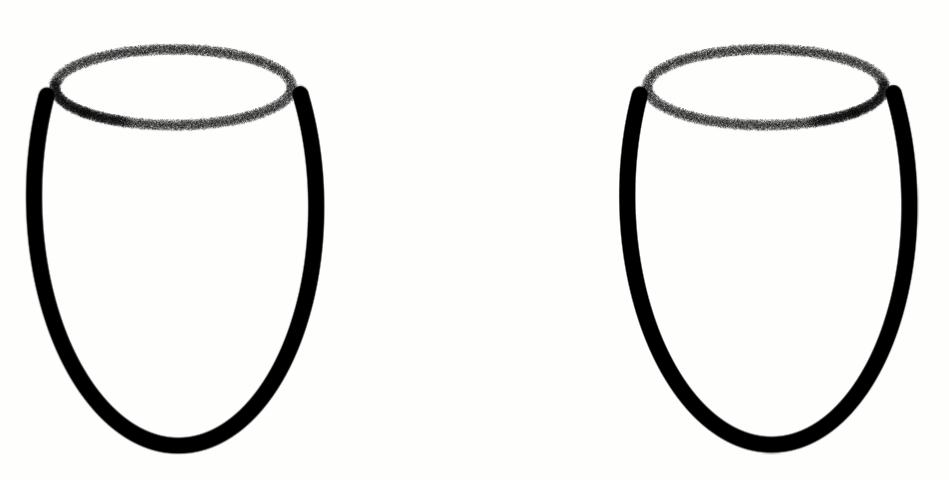}
\end{array}
\qquad \text{and} \qquad
\begin{array}{c}
\includegraphics[width=0.12\linewidth]{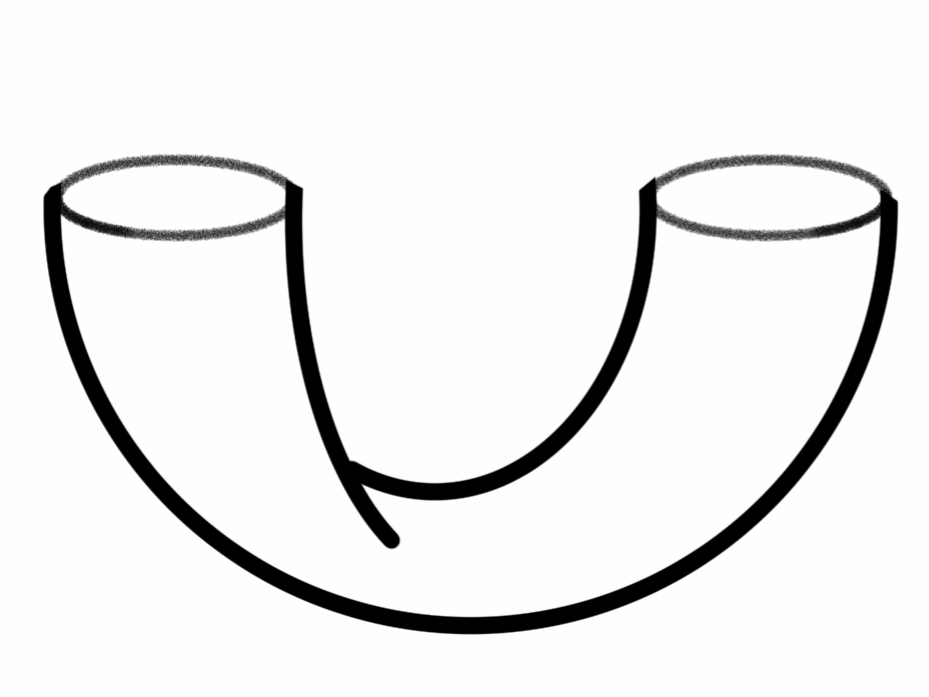}
\end{array}\,.
\label{states}
\ee
The first state is quite obviously separable and hence non-entangled. The second state should be entangled, as the picture suggests. Let us compute the von Neumann entropy of both states to confirm this.

First, we need the corresponding density matrices. These are computed by definition $\rho=|\psi\rangle\langle\psi|$, which in the TQFT language means taking
$$\hat{\rho}_1 \ = \ \begin{array}{c}
 \includegraphics[width=0.2\linewidth,clip=true,trim=400pt 0pt 300pt 0pt]{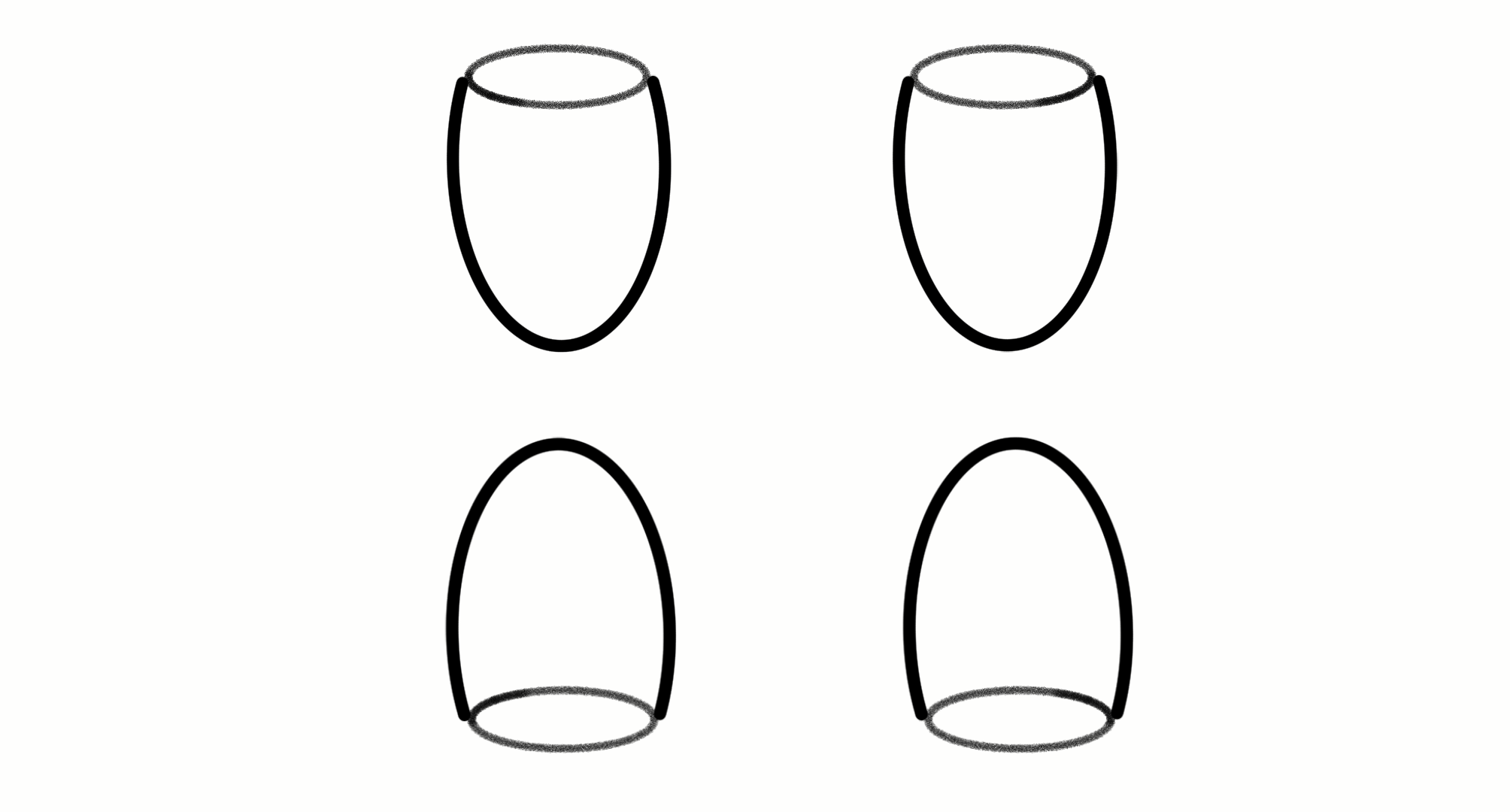}
 % cob1.png: 0x0 pixel, 0dpi, 0.00x0.00 cm, bb=
 \end{array}\,, \qquad
\hat{\rho}_2 \ = \
\begin{array}{c}
 \includegraphics[width=0.2\linewidth,clip=true,trim=400pt 0pt 300pt 0pt]{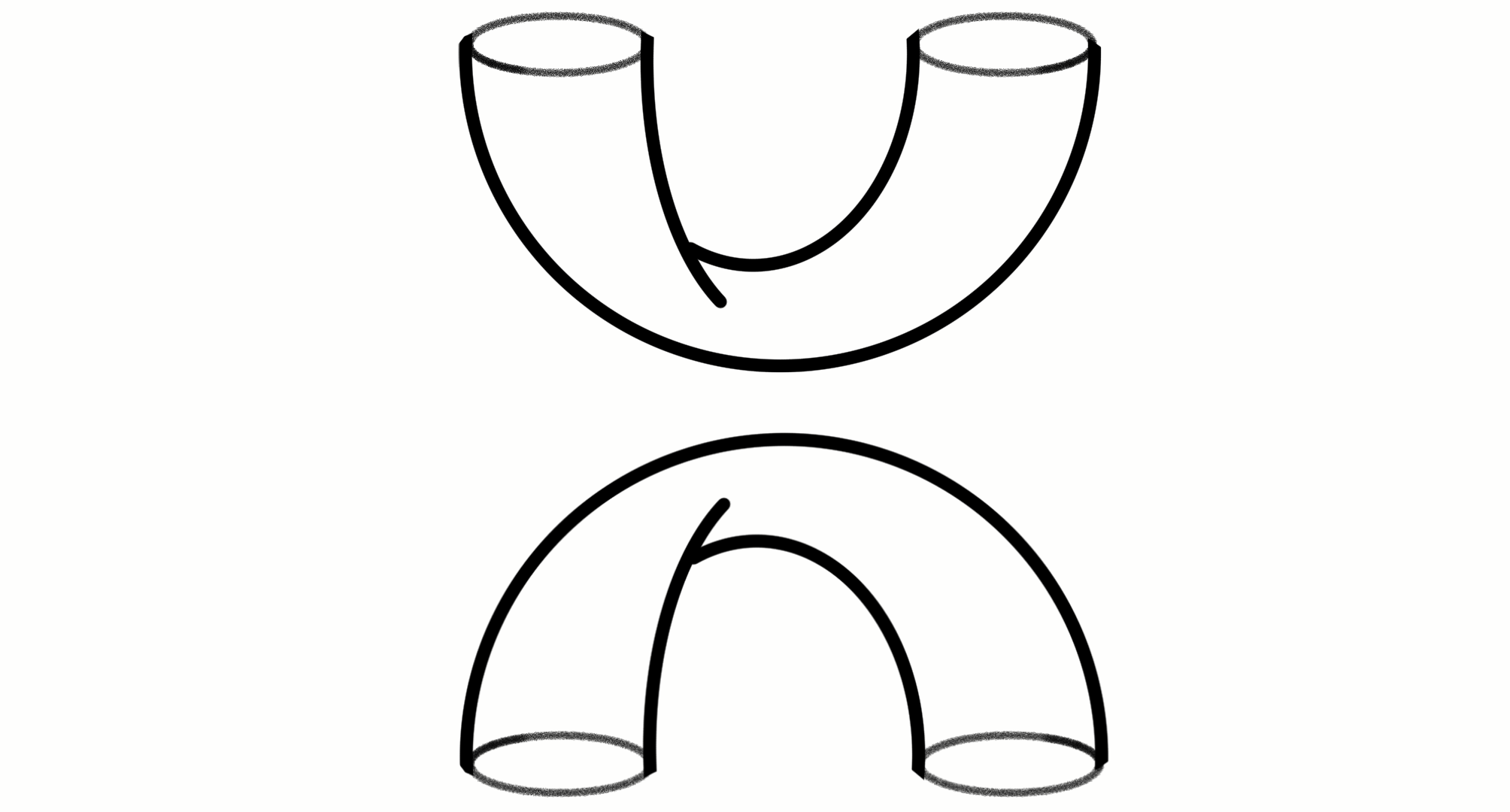}
\end{array}\,.$$
The density matrices should be normalized to have unit trace. The normalization factors are
$$\Tr\hat{\rho}_1 \ = \ \left[\begin{array}{c}
 \includegraphics[width=0.07\linewidth]{./sphere.png}
 % cob1.png: 0x0 pixel, 0dpi, 0.00x0.00 cm, bb=
 \end{array}\right]^2\,, \qquad
\Tr\hat{\rho}_2 \ = \
\begin{array}{c}
 \includegraphics[width=0.1\linewidth,clip=true,trim=400pt 0pt 300pt 0pt]{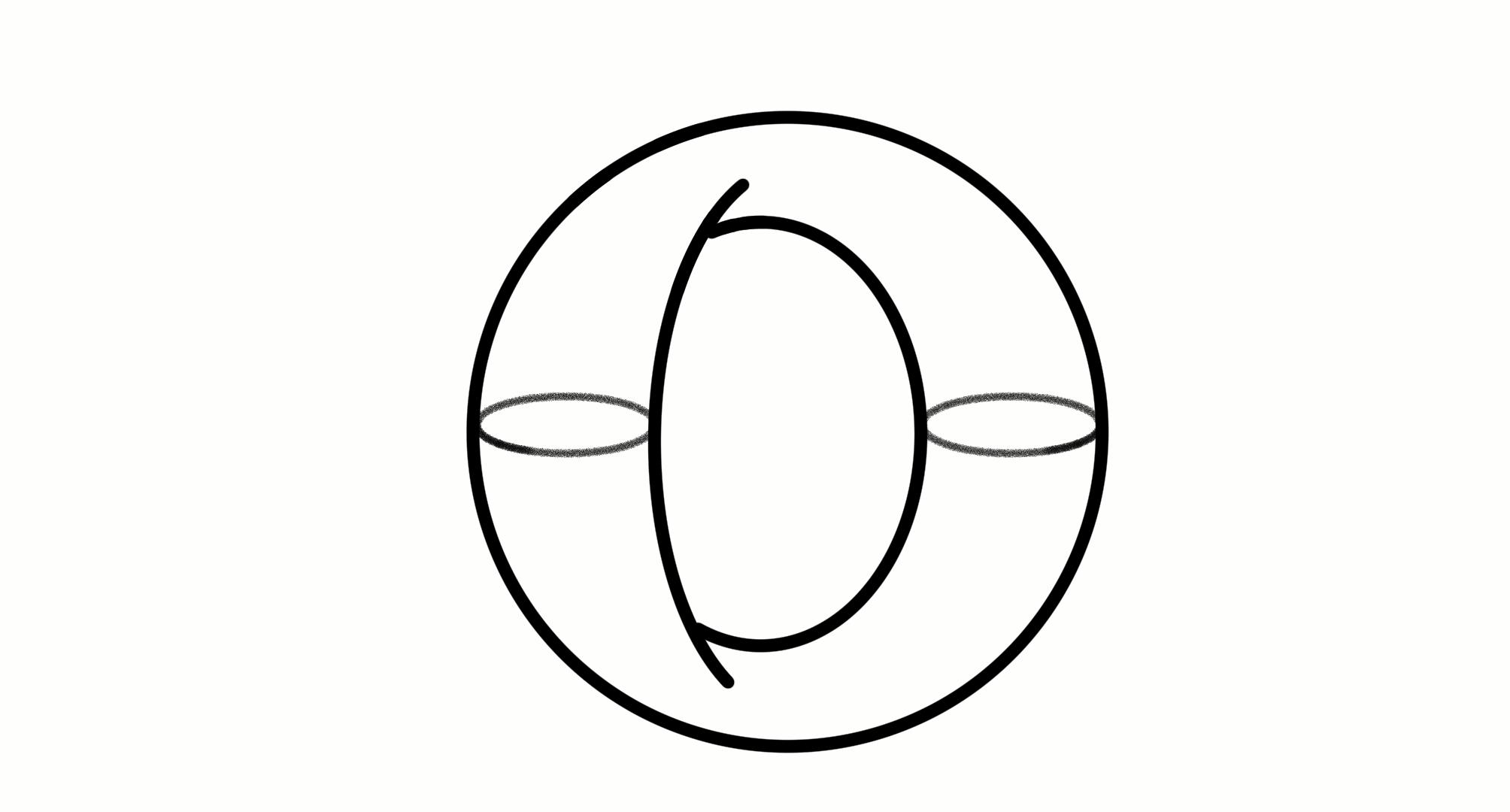}
\end{array}\,.$$
Here traces are computed by gluing together every top circle with the corresponding bottom circle. The result of such gluing is a manifold with empty boundary, a $C$-number.

The von Neumann entropy is defined in terms of a reduced density matrix. Let us trace out the right circles in both examples. Obviously, the reduced density matrices of the left subsystem would be
\be
\label{rdms}
\rho_{1L}  \equiv  \tr\!_R \rho_1 =      \left[
\begin{array}{c}\includegraphics[width=0.07\linewidth]{./sphere.png}\end{array}
\right]^{-1} \times
\begin{array}{c}\includegraphics[width=0.08\linewidth,clip=true,trim=550pt 0pt 850pt 0pt]{./rho1.png}\end{array}
\quad  \rho_{2L}  \equiv  \tr\!_R \rho_2  =      \left[
\begin{array}{c}\includegraphics[width=0.1\linewidth,clip=true,trim=550pt 70pt 450pt 90pt]{./zs2s1.png}\end{array}
\right]^{-1} \times
\begin{array}{c}\includegraphics[width=0.08\linewidth,clip=true,trim=600pt 0pt 900pt 100pt]{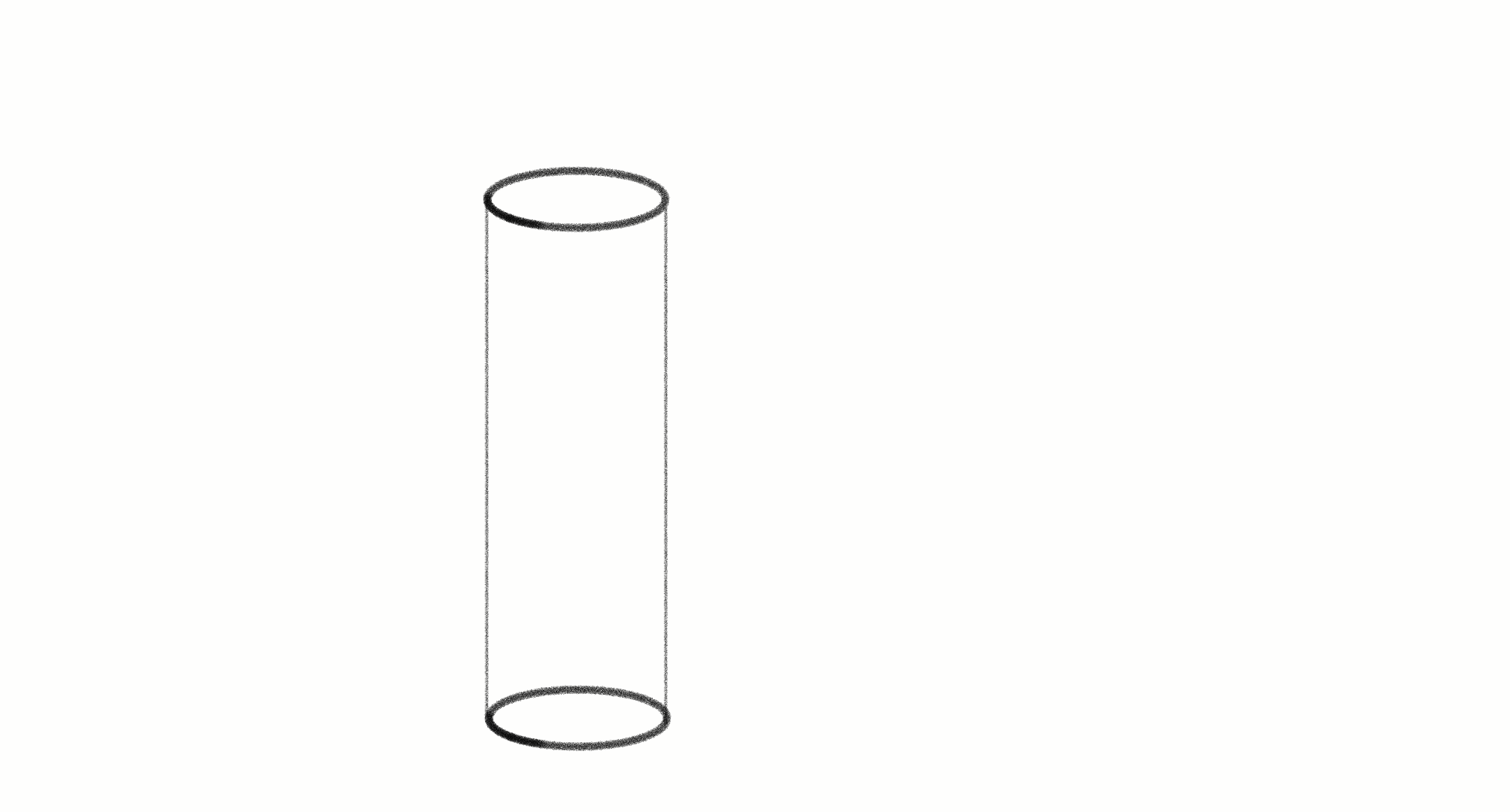}\end{array}
\ee

Although it is not immediately clear how to directly apply the von Neumann formula
$$
S_{L} \ = \ - \tr\!_L\, \rho_L \log\rho_L\,,
$$
it is straightforward to apply the replica trick~\cite{Dong:08}, first computing $\tr\!_L\,\rho_L^n$, then analytically continuing in $n$ and evaluating
$$
S_{L} \ = \ -\left. \frac{d}{dn}\tr\!_L\,\rho_L^n\right|_{n=1}.
$$
Following the above discussion, the evaluation of $\rho^n$ includes stacking $n$ copies of the reduced density matrices~(\ref{rdms}) on top of each other, sewing them together and counting the normalization factors. It is clear that any power of $\rho_L$ will be proportional to the original matrix, and the only difference will be in the accumulated normalization factors. It is then straightforward to find
\be
\label{entropy}
S_L(\rho_1) \ = \  0\,, \qquad S_L(\rho_2) \ = \
\log \left[
\begin{array}{c}\includegraphics[width=0.1\linewidth,clip=true,trim=550pt 70pt 450pt 90pt]{./zs2s1.png}\end{array}
\right].
\ee
As expected, the entropy of a separable state is zero. To understand the meaning of the expression for the entropy of the state with two cobordant subsystems we need to give some details about the symbol appearing in the logarithm. We will explain this by considering some examples below.

Before proceeding, we comment on the operator, which appears in the right density matrix in equation~(\ref{rdms}) (the cylinder). This can be thought as of a mixed density matrix, as opposed to the separable reduced density matrix on the left. The equivalence between the operator (cylinder) and the entangled pair in equation~(\ref{states}) is even more obvious in the TQFT presentation.

Finally, one can imagine that an entangling operator can have a presentation
$$
\begin{array}{c}
\includegraphics[width=0.1\linewidth]{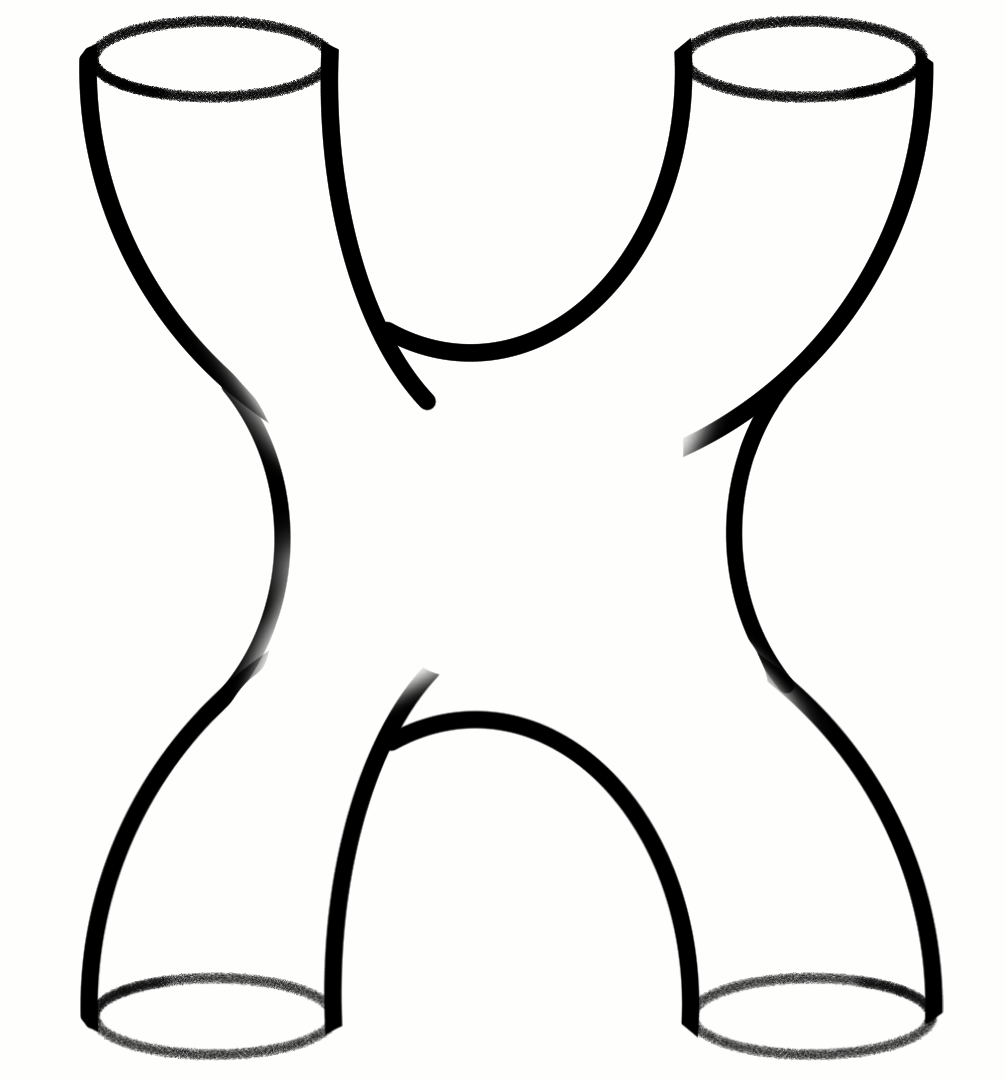}
\end{array}\,.
$$

\section{Entanglement and Wilson lines}
\label{sec:examples}

Let us explain result~(\ref{entropy}) of the graphical calculation from the previous section. First, one can connect it to the results of~\cite{Balasubramanian:2016sro}, where the entanglement of links was introduced. Namely, consider $\Sigma$ to be a two dimensional torus $T^2$. It is known that in case of $T^2$ one can associate to it a family of Hilbert spaces $\Hc_N^k$, labeled by integers $N\geq 2$, $k\geq 0$ of dimension\footnote{These Hilbert spaces are in one-to-one correspondence with fusion algebra of $SU(N)_k$ WZNW theories. Alternatively, they can be defined through the functional integral of level $k$ $SU(N)$ Chern-Simons theory~\cite{Witten:1988hf}.}
$$\dim \Hc_N^k \ = \ \left(\begin{array}{c}
k+N-1 \\
N-1
                            \end{array}
\right)\,.$$
An entangled pair, in $\Sigma=T^2$ case can be introduced by a solid torus with a hollow tube inside:
\be
\label{torpsi}
|\psi\rangle \ = \ \begin{array}{c}
\includegraphics[width=0.2\linewidth]{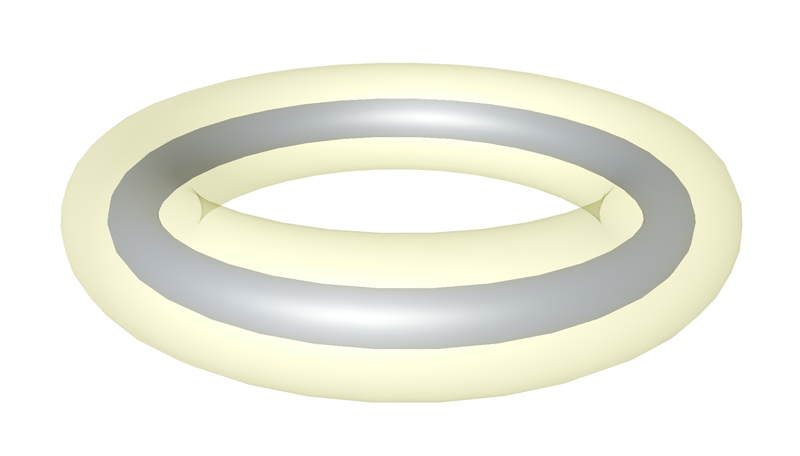}
                   \end{array}\,.
\ee
The same figure represents the unnormalized reduced density matrix. Following the steps of the previous section we arrive at equation~(\ref{entropy}), where the donut denotes the map $Z:T^2\times S^1\to C$. For $Z(T^2)=\Hc_N^k$, the result of this map is nothing but~\cite{Witten:1988hf}
\be
S_L \ = \ \log Z(T^2\times S^1) \ = \ \log\dim \Hc_N^k\,.
\ee
This is consistent with reference~\cite{Balasubramanian:2016sro}. It is not hard to see that, in their language, $|\psi\rangle$ is a state associated to a Hopf link, which is maximally entangled.

One can also consider the case of the two-dimensional sphere $\Sigma=S^2$. Again, there are Hilbert spaces that can be associated with $S^2$. However, for the proper $S^2$, the Hilbert space is one-dimensional. In order to have a non-trivial example, one needs to consider spheres with at least three marked points (punctures). The minimum number of punctures is three, then the dimension of the Hilbert space is determined by numbers
$$
\dim\Hc_\Sigma^3 \ = \ N_{ab}^c\,,
$$
which can be found from the fusion algebra of $SU(N)_k$ WZNW model. These numbers tell how many ways fields of colors ($su(N)_k$ representations) $a$ and $b$ can fuse to make the field of color $c$. In a generic situation these numbers are either zero, or one, so it is more convenient to consider examples with at least four punctures.

Riemann surfaces with punctures are equivalent to surfaces with boundaries. In the TQFT construction one should, consequently, consider cobordisms of such surfaces. In this case one should first introduce the cobordism of the boundaries: $\partial\Sigma = \partial m$ (the point is that interesting $m$ is not the same as $\Sigma$ itself), and then define the full cobordism $M$, with $\partial M = \Sigma - m$. In particular, surfaces with punctures can be cobordant to other surfaces with punctures, if punctures are themselves cobordant -- and the ``interesting" case is when punctures are connected by strings, or Wilson lines in terms of the explicit Chern-Simons presentation of the TQFT functor.

We now consider examples of states
\be
\begin{array}{c}
\includegraphics[width=0.15\linewidth]{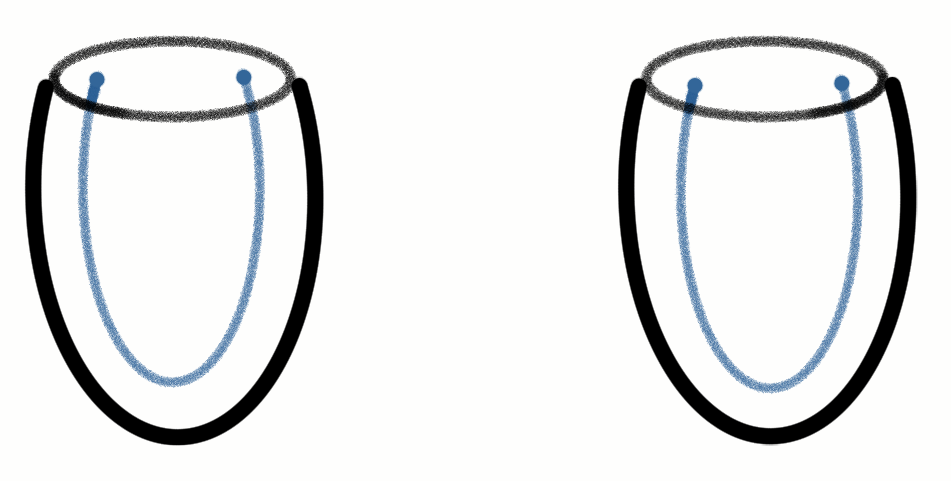}
\end{array}
\qquad \text{and} \qquad
\begin{array}{c}
\includegraphics[width=0.17\linewidth]{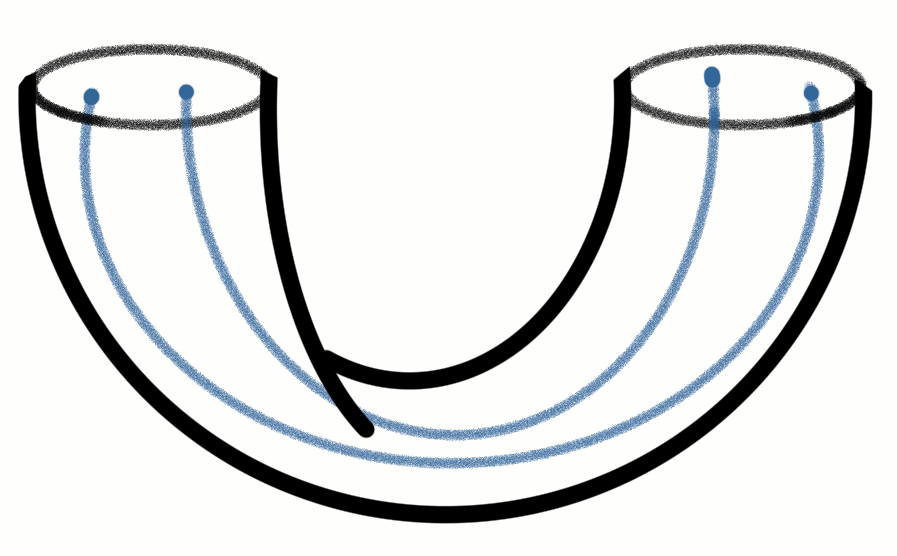}
\end{array}\,,
\label{wlines}
\ee
threaded by Wilson lines.

The calculation of the entanglement entropy will be somewhat complicated by the presence of the Wilson lines, since one needs to keep track of what happens with them when the density matrix is reduced and replicated. We discuss few special cases.

\begin{itemize}

\item Start with a pair of cobordant $\Sigma_1$ and $\Sigma_2$, with Wilson lines only connecting punctures locally, as in the diagram
\be
\label{psi0}
|\Psi_0\rangle\ = \ \begin{array}{c}
\includegraphics[width=0.15\linewidth]{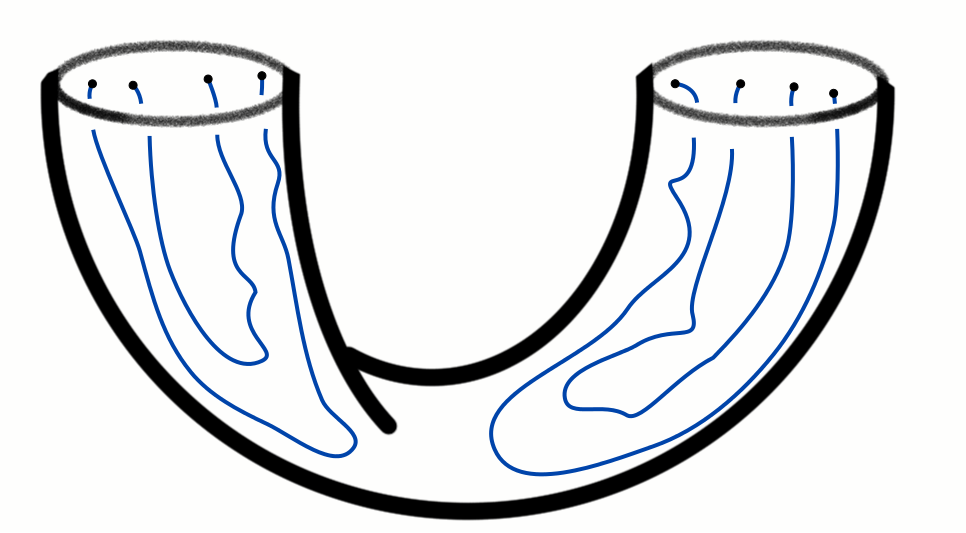}
\end{array}
\qquad \Longrightarrow \qquad
\rho_L \ \propto \  \begin{array}{c}
\includegraphics[width=0.05\linewidth]{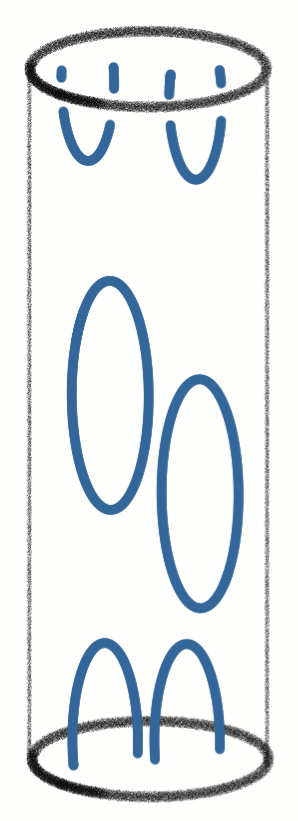}
\end{array}
\ee
Following the steps of the derivation of equation~(\ref{entropy}) the trace of the reduced density matrix raised to the power $n$ will be expressed in terms of the link invariant of unlinked circles in $S^2\times S^1$ that do not wind around $S^1$. The invariants of such configurations always factorize in product of invariants of a single circle (unknot). Consequently, properly normalized trace is always $n$-independent and the entropy is zero.

\item Now consider Wilson lines that extend from $\Sigma_1$ to $\Sigma_2$ without braiding, as in the right example of equation~(\ref{wlines}). The entanglement entropy computed by equation~(\ref{entropy}) in this case will be
\be
S_L \ = \ \log Z(S^2\times S^1;a_1,\ldots,a_n) \ = \ \log \dim \Hc_{S^2}^n\,,
\ee
where $a_i$ label Wilson lines of different color threading the $S^2\times S^1$ along the $S^1$. The entropy is the logarithm of the dimension of the Hilbert space associated with $S^2$ with $n$ punctures. With zero, one or two punctures such Hilbert space is at most one-dimensional, but for $n\geq 3$ one can easily find cases with $\dim \Hc_{S^2}^n>1$, which correspond to maximal entanglement.

 Our prime example of a maximally entangled pair is a pair of cobordant spheres with four punctures connected in such a way that every Wilson line is extended between the two spheres.
\be
\label{psi}
|\Psi\rangle \ = \ \begin{array}{c}
\includegraphics[width=0.15\linewidth]{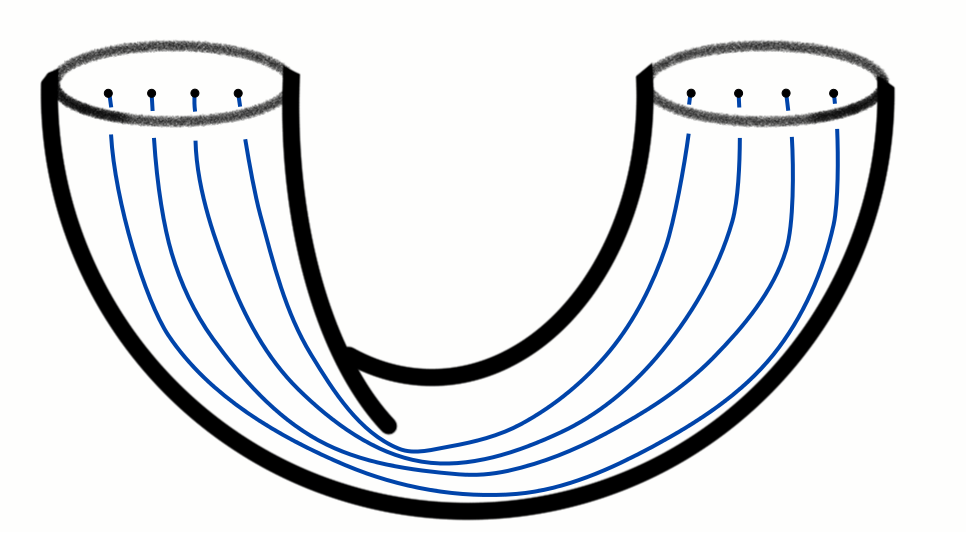}
\end{array}
\qquad \Longrightarrow \qquad
\rho_L \ \propto \  \begin{array}{c}
\includegraphics[width=0.07\linewidth]{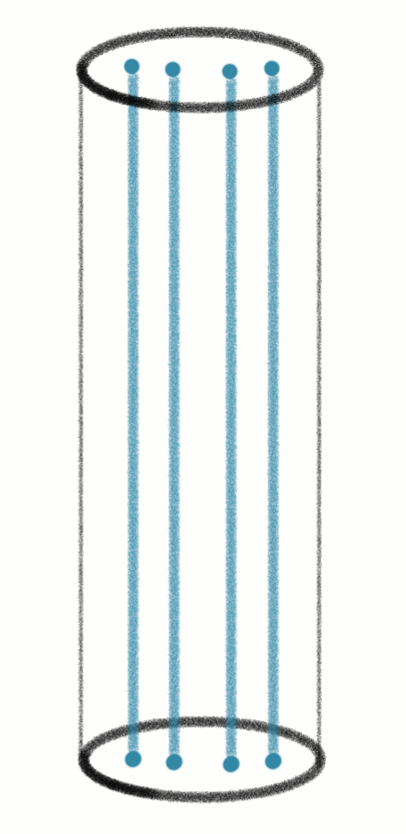}
\end{array}
\,.
\ee

\item We note that any braiding of Wilson lines extended between two spheres in states~(\ref{psi0}) or (\ref{psi})   does not affect the entropy, since it amounts to a local (diffeomorphism) operation on either of the spheres, of which the von Neumann entropy is independent. This is the standard property of the von Neumann entropy under local unitary transformations: let $|\Psi\rangle = |\psi_1\rangle \otimes |\psi_2\rangle$. Apply a transformation $|\psi_2\rangle\to U|\psi_2\rangle$. Density matrix transforms as $\rho\to U\rho U^{\dagger}$. Since the entropy is defined through trace of the density matrix and its products, local unitaries do not affect the entropy. This can also be understood in terms of the braiding of the Wilson lines.

\item A transformation acting non-trivially on the entropy is the non-local braiding that might take $|\Psi\rangle$ in equation~(\ref{psi}) to~(\ref{psi0}), or to an intermediate situation like
\be
|\Psi'\rangle  =  \begin{array}{c}
\includegraphics[width=0.15\linewidth]{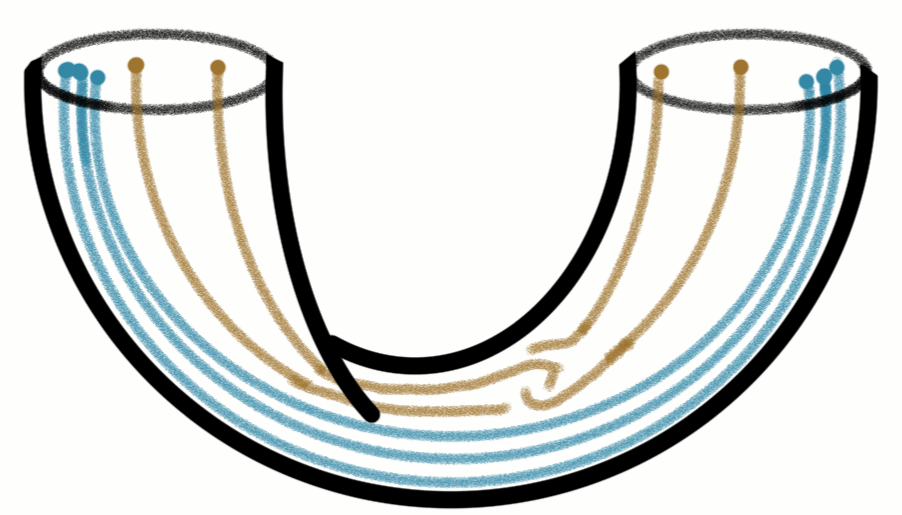}
\end{array}
\quad \Longrightarrow \qquad
\rho_L \ \propto  \begin{array}{c}
\includegraphics[width=0.05\linewidth]{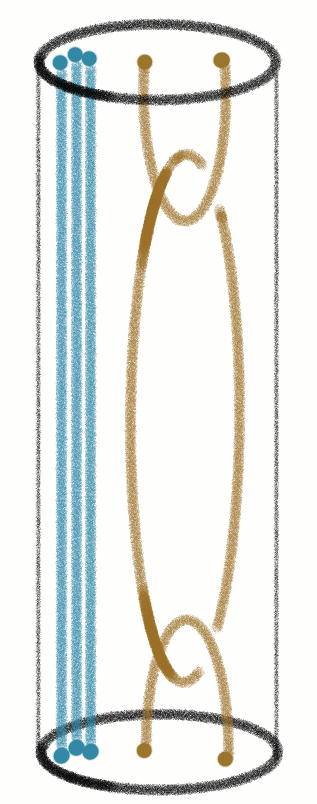}
\end{array}
\quad \Longrightarrow \qquad
\tr \rho_L^n \ \propto   \begin{array}{c}
\includegraphics[width=0.15\linewidth]{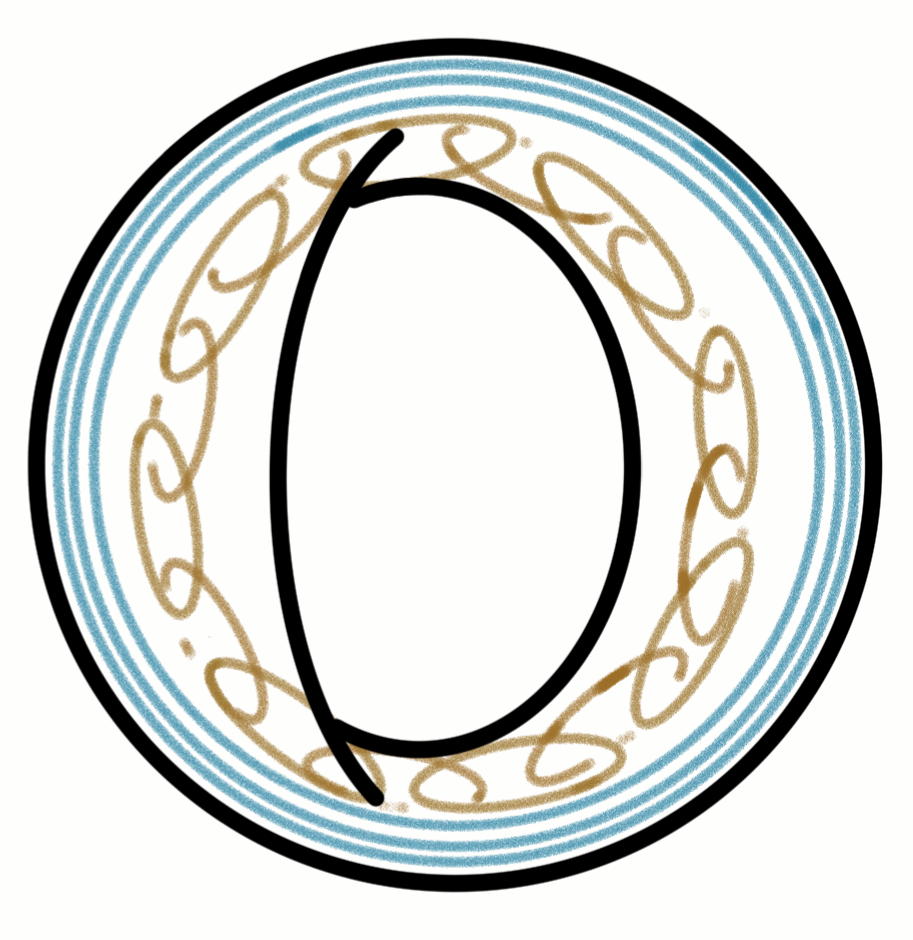}
\end{array}
\label{chained}
\ee
In the latter case, in principle, one finds an example of a non-trivial invariant: one needs to calculate an $S^2\times S^1$ invariant of chains with $n$ segments, with possible additional lines. The simplest cases of $S^3$ invariants of such \emph{necklaces} have been recently studied in Ref.~\cite{MMM}.

\end{itemize}

\section{Conclusions}
\label{sec:conclusions}

In the above study we introduced the idea of entanglement between subsystems in TQFT as cobordisms between topological spaces $\Sigma$, representing those parts. We checked that this gives a necessary, but not sufficient condition for entanglement. In~\cite{Balasubramanian:2016sro} the sufficient condition was a non-trivial linking of $\Sigma_i$, which in example~(\ref{torpsi}) above meant that the hollow tube inside the solid torus should wind the non-contractible cycle. For $\Sigma=S^2$, entanglement requires a sufficient number of Wilson lines to connect the entangled subsystems. The necessary condition of the entanglement is thus keeping the Hilbert space notrivial at every section of the cobordism.

Examples considered in section~\ref{sec:examples} show that states like~(\ref{psi}), with all Wilson lines extending from $\Sigma_1$ to $\Sigma_2$ are, in principle, maximally entangled, while the states like~(\ref{psi0}), which are only locally weaved by Wilson lines are not entangled at all. Indeed, in the latter case the Wilson lines annihilate each other before reaching the other part of the system. The reduced density matrix is purified. More generally, Wilson lines do not have to go all the way towards the other boundary. They can meet and braid (and entangle) with Wilson lines that come from the other boundary, as in state~(\ref{chained}).

Interestingly, in both $T^2$ and $S^2$ examples, maximal entanglement is achieved on configurations with a simpler topology (braiding). In the case of $T^2$, it is achieved on the Hopf link, which is the simplest non-trivial linking of two circles.  In the case of $S^2$, state~(\ref{psi}) with lines connecting opposite boundaries, is already maximally entangled. The entropy is insensitive to any local braiding, while non-local braiding, in general, reduces the entropy.

We notice that examples considered here provide a different realization of the Aravind's proposal~\cite{Aravind}, which to the authors' knowledge,  was the first attempt to connect the quantum and topological entanglements. It was suggested, in particular, to view the Hopf link as an EPR, or Bell pair, while the Borromean rings were proposed as an example of a GHZ state: removing a single link leaves the remaining pair unlinked.

We would also like to mention an apparent connection of the ideas presented here to the $ER=EPR$ discussion of Ref.~\cite{Maldacena:2013xja}. The main suggestion of the $ER=EPR$ equivalence, that a two-sided black hole, connected by the Einstein-Rosen (ER) bridge is equivalent to the entangled EPR pair, can be cast in the language of the present paper. The ER bridge realizes a cobordism between two asymptotic boundaries of the black hole (see also Ref.~\cite{Gharibyan:2013aha}). While this paper was being prepared, a paper discussing entanglement of cobordant regions in AdS/CFT appeared~\cite{VanRaamsdonk:2018zws} based on the ideas of Ref.~\cite{Balasubramanian:2014hda}.

Finally, we repeat our claim in~\cite{topquco} that, like quantum programming, the whole subject of quantum entanglement can appear reducible to the theory of braids and knots. In this context it deserves mentioning that a powerful formulation of knot theory in Ref.~\cite{Khovanov} involves singular  (one-dimensional) spaces $\Sigma$ and singular cobordisms (foams) between them, but in many respects looks surprisingly similar to what we discussed above.

\paragraph{Acknowledgements.} DM would like to thank P.~Kaputa and F.~Toppan for useful discussions and for communicating works~\cite{Dong:08} and~\cite{Kauffman:2013}. DM would also like to thank H.~Gharibyan for interesting discussions on the connections to Ref.~\cite{Maldacena:2013xja} and J.~Barreto for the help with graphics. This work was supported by Russian Science Foundation grant No~18-71-10073.

\end{document}